\documentclass[aip]{revtex4-1}
\usepackage{graphicx}

\draft

\begin{document}

\title{Magnetic cryocooling with Gd$^{3+}$ centers in a light and compact framework}

\author{G. Lorusso}
\affiliation{Instituto de Ciencia de Materiales de Arag\'{o}n (ICMA), CSIC $-$ Universidad de Zaragoza, Departamento de F\'{i}sica de la Materia Condensada, 50009 Zaragoza, Spain}

\author{J. W. Sharples}
\affiliation{School of Chemistry and Photon Science Institute, The University of Manchester, M13-9PL Manchester, United Kingdom}

\author{E. Palacios}
\affiliation{Instituto de Ciencia de Materiales de Arag\'{o}n (ICMA), CSIC $-$ Universidad de Zaragoza, Departamento de F\'{i}sica de la Materia Condensada, 50009 Zaragoza, Spain}

\author{O. Roubeau}
\affiliation{Instituto de Ciencia de Materiales de Arag\'{o}n (ICMA), CSIC $-$ Universidad de Zaragoza, Departamento de F\'{i}sica de la Materia Condensada, 50009 Zaragoza, Spain}

\author{E. K. Brechin}
\affiliation{School of Chemistry, The University of Edinburgh, EH9-3JJ Edinburgh, United Kingdom}

\author{R. Sessoli}
\affiliation{Department of Chemistry and INSTM, Universit\`{a} degli Studi di Firenze, 50019 Sesto Fiorentino, Italy}

\author{A. Rossin}
\affiliation{Istituto di Chimica dei Composti Organometallici (ICCOM), CNR, 50019 Sesto Fiorentino, Italy}

\author{F. Tuna}
\affiliation{School of Chemistry and Photon Science Institute, The University of Manchester, M13-9PL Manchester, United Kingdom}

\author{E. J. L. McInnes}
\affiliation{School of Chemistry and Photon Science Institute, The University of Manchester, M13-9PL Manchester, United Kingdom}

\author{D. Collison}
\affiliation{School of Chemistry and Photon Science Institute, The University of Manchester, M13-9PL Manchester, United Kingdom}

\author{M. Evangelisti}
\homepage[]{http://molchip.unizar.es/}
\affiliation{Instituto de Ciencia de Materiales de Arag\'{o}n (ICMA), CSIC $-$ Universidad de Zaragoza, Departamento de F\'{i}sica de la Materia Condensada, 50009 Zaragoza, Spain}

\date{\today}

\begin{abstract}
The magnetocaloric effect of gadolinium formate, Gd(OOCH)$_3$, is experimentally determined down to sub-Kelvin temperatures by direct and indirect methods. This 3D metal-organic framework material is characterized by a relatively compact crystal lattice of weakly interacting Gd$^{3+}$ spin centers interconnected via light formate ligands, overall providing a remarkably large magnetic:non-magnetic elemental weight ratio. The resulting volumetric magnetic entropy change is decidedly superior in Gd(OOCH)$_3$ than in the best known magnetic refrigerant materials for liquid-helium temperatures and low-moderate applied fields.
\end{abstract}


\maketitle

Recent years have witnessed a terrific increase in the number of molecule-based materials proposed as magnetic refrigerants for liquid-helium temperatures.~\cite{Evangelisti10,Sessoli12,Torres00,Spichkin01,Affronte04,Evangelisti05,Evangelisti09,Evangelisti11,Martinez12,Manuel06,Evangelisti07,Sedlakova09,Sibille12,Guo12,Lorusso12} Refrigeration proceeds adiabatically via the magnetocaloric effect (MCE), which describes the changes of magnetic entropy $(\Delta S_m)$ and adiabatic temperature $(\Delta T_{ad})$, following a change in the applied magnetic field $(\Delta B)$. As in the first paramagnetic salt that permitted sub-Kelvin temperatures to be reached in 1933,~\cite{Giauque33} gadolinium is often present because its orbital angular momentum is zero and it has the largest entropy per single ion.~\cite{Evangelisti10} How to spatially assemble the Gd$^{3+}$ spin centers is vital for designing the ideal magnetic refrigerant.~\cite{Martinez12,Lorusso12} On the one hand, the magnetic density should be maximized, for instance, by limiting the amount of non-magnetic elements which act passively in the physical process. On the other hand, the magnetic ordering for $B=0$ should not develop, causing the decrease of MCE, above the target working temperature of the refrigerant. Therefore a compromise becomes necessary, especially for reaching very low temperatures.

This letter focuses on gadolinium formate, whose chemical formula reads Gd(OOCH)$_3$, which belongs to the class of metal-organic framework (MOF) materials. Other three-dimensional MOF materials have recently attracted the interest for their cooling properties, combined with their synthetic variety and intrinsic robustness.~\cite{Manuel06,Evangelisti07,Sedlakova09,Sibille12,Guo12,Lorusso12} We shall see below that, while presenting a sub-Kelvin ordering temperature, Gd(OOCH)$_3$ is also characterized by a relatively high packing density of Gd$^{3+}$ ions, linked only to light formate ligands. The resulting MCE, that we estimate by direct and indirect methods, sets Gd(OOCH)$_3$ in an enviable position within this research area.

\begin{figure}[b!]
\centering\includegraphics[angle=0,width=9cm]{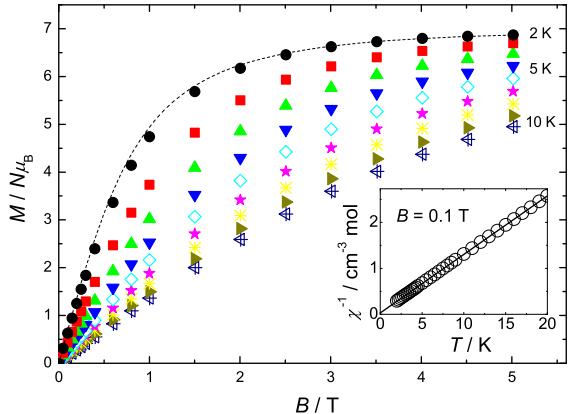}
\caption{Field-dependence of the experimental molar magnetization $M$ for temperatures ranging from 2 to 10~K, with a 1~K step between adjacent isothermal curves. Dashed line is the calculated $M$ of a paramagnet for $s=7/2$, $g=2$ and $T=2$~K. Inset: temperature-dependence of the inverse of the experimental molar susceptibility $\chi$ collected for $B=0.1$~T and Curie-Weiss fit (solid line).}
\end{figure}

Single-crystal structure determination of Gd(OOCH)$_3$ completes the original results derived from powder diffraction.~\cite{Pabst43,Brechin12} No previous magnetic measurements on Gd(OOCH)$_3$ are reported in the literature, except for initial M\"{o}ssbauer experiments.~\cite{Cashion73} Magnetization measurements down to 2~K and heat capacity measurements using the relaxation method down to $\approx 0.35$~K were carried out on powder samples by means of commercial setups for $0<B<5$~T and $0<B<7$~T, respectively. Direct measurements of the MCE were performed on a powder sample using a dedicated thermal sensor, installed in the same setup employed for the heat capacity experiments.

Figure~1 shows the measured molar magnetization $M$ for temperatures within $2-10$~K. The magnetization saturates to the expected value of 7~$\mu_{\rm B}$ for a Gd$^{3+}$ spin moment, according to which $s=7/2$ and $g=2$. The $M(T)$ curves can be well described by a Brillouin function -- see, e.g., the dashed line in Fig.~1 for an ideal paramagnet at $T=2$~K. Deviations of the experimental data from the paramagnetic behavior are barely noticeable only for the lowest temperatures, and can be ascribed to the presence of a weak antiferromagnetic interaction. This is corroborated by the $T$-dependence of the magnetic susceptibility $\chi$. As shown by the solid line in the inset of Fig.~1, the susceptibility data can be fitted above 2~K to a Curie-Weiss law $\chi=g^{2}\mu_{\rm B}^{2}s(s+1)/[3k_{\rm B}(T-\theta)]$, obtaining a negative, though small, $\theta=-0.3$~K, which suggests that the Gd$^{3+}$ moments are weakly antiferromagnetically correlated in the paramagnetic phase.

\begin{figure}
\centering\includegraphics[angle=0,width=8cm]{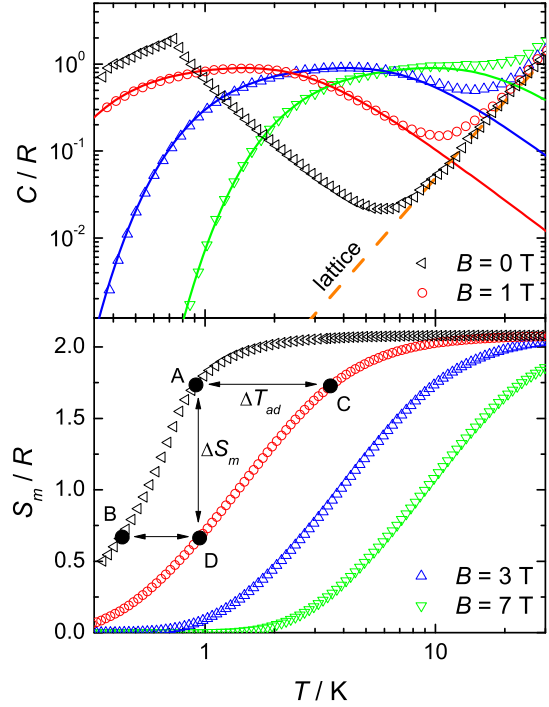}
\caption{Top: temperature-dependence of the heat capacity $C$ normalized to the gas constant $R$ collected for $B=0,1,3$, and 7~T, as labeled. Solid thick lines are the calculated Schottky contributions for the corresponding $B$, and dashed line is the fitted lattice contribution. Bottom: $T$-dependence of the experimental magnetic entropy $S_{m}$ normalized to the gas constant $R$ for several $B$, as obtained from the magnetic contribution $C_{m}$ to the total heat capacity. Highlighted examples of magnetic entropy change $\Delta S_{m}$ between states ${\rm A}\leftrightarrow{\rm D}$ and adiabatic temperature changes $\Delta T_{ad}$ between states ${\rm A}\leftrightarrow{\rm C}$ and ${\rm B}\leftrightarrow{\rm D}$.}
\end{figure}

The top panel of Figure~2 shows the measured low-temperature heat capacity $C$, normalized to the gas constant $R$, as a function of temperature for several applied fields. A sharp lambda-like peak can be observed in the zero-field data for $T_{C1}\simeq 0.8$~K, denoting the presence of a phase transition, which is accompanied by a smooth and tiny feature at $T_{C2}\approx 0.4$~K. The magnetic origin of both anomalies is proved by the fact that external applied fields quickly and fully suppress them.~\cite{note} In agreement with $M(T,B)$, the analysis of the field-dependent $C$ reveals that magnetic interactions between the Gd$^{3+}$ spin centers are relatively weak, since applied fields $B\geq 1$~T are sufficient for fully decoupling all spins. As indeed shown in Fig.~2, the calculated Schottky contributions (solid lines) for the field-split levels of the non-interacting $s=7/2$ multiplet nicely account for the magnetic contribution $C_{m}$ to the experimental heat capacity. For $T\gtrsim 7$~K, a large field-independent contribution appears, which can be attributed to the lattice phonon modes of the crystal. The dashed line in the top panel of Fig.~2 represents a fit to this contribution, with the well-known Debye function yielding a value of $\Theta_{\rm D}=168$~K for the Debye temperature, which is remarkably large among molecular~\cite{Evangelisti06} and MOF~\cite{Lorusso12} materials, denoting a relatively stiff lattice. A larger $\Theta_{\rm D}$ implies a correspondingly lower lattice entropy in the low-$T$ region, ultimately favoring the MCE. From the experimental heat capacity the temperature dependence of the magnetic entropy $S_{m}(T)$ is derived by integration, i.e.,
\begin{equation}\label{eqS}
S_{m}(T)=\int_{0}^{T}\frac{C_{m}(T)}{T}{\rm d}T,
\end{equation}
where $C_{m}$ is obtained by subtracting the lattice contribution to the total $C$ measured. The so-obtained $S_{m}(T)$ is shown in the bottom panel of Fig.~2 for the corresponding applied fields. For $B=0$, the lack of experimental $C_{m}$ for $T\lesssim 0.3$~K has been taken into account by matching the limiting $S_{m}$ at high $T$ with the value obtained from the in-field data. One can notice that there is a full entropy content of $R{\rm ln}(8)$ per mole Gd$^{3+}$ involved, as expected from $R{\rm ln}(2s+1)$ and $s=7/2$.

\begin{figure}[t!]
\centering\includegraphics[angle=0,width=8cm]{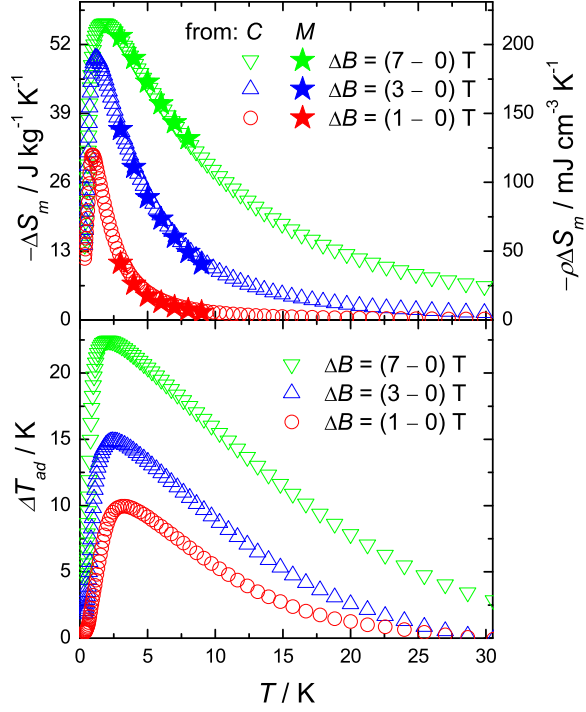}
\caption{Top: temperature-dependence of the magnetic entropy change $\Delta S_{m}$, as obtained from magnetization and heat capacity data (Figs.~2 and 3, resp.) for the indicated applied-field changes $\Delta B$. Vertical axis reports units in J~kg$^{-1}$~K$^{-1}$ (left) and volumetric mJ~cm$^{-3}$~K$^{-1}$ (right). Bottom: $T$-dependence of the adiabatic temperature change $\Delta T_{ad}$, as obtained from heat capacity data for the indicated $\Delta B$.}
\end{figure}

Next, we {\it indirectly} evaluate the MCE of Gd(OOCH)$_3$ from the experimental data presented so far. From the bottom panel of Fig.~2, we straightforwardly obtain the magnetic entropy changes $\Delta S_{m}(T,\Delta B)$ for different applied field changes $\Delta B=B_{f}-B_{i}$. The so-obtained results are depicted in Figure~3. A similar set of data can also be derived from an isothermal process of magnetization by employing the Maxwell relation, i.e., $\Delta S_{m}(T,\Delta B)=\int_{B_{i}}^{B_{f}}[\partial M(T,B)/\partial T]_{B}{\rm d}B$. From the experimental $M(T,B)$ data in Fig.~1, we then obtain curves that rather beautifully corroborate the corresponding results previously derived from heat capacity -- see top panel of Fig.~3. Furthermore, to a cooling process under adiabatic conditions, one naturally associates a temperature change whose estimate is made feasible by knowing $C$ and thus $S_{m}$. The bottom panel of Fig.~3 shows $\Delta T_{ad}(T,\Delta B)$, where $T$ denotes the final temperature of the adiabatic cooling, e.g., for going from states C$(T=3.4~{\rm K},B=1~{\rm T})$ to A$(T=0.95~{\rm K},B=0)$ in Fig.~2.

\begin{figure}[t!]
\centering\includegraphics[angle=0,width=8cm]{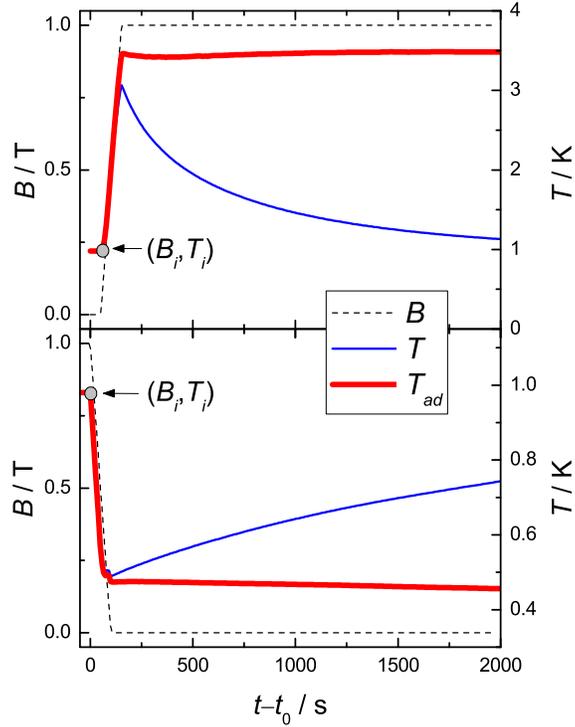}
\caption{Time evolution of the applied field $B$, experimental temperature $T$ and deduced adiabatic temperature $T_{ad}$, as labeled, during a magnetization (top) and a demagnetization (bottom) process, both starting from $B_i$ and $T_i=0.95$~K.}
\end{figure}

A far more elegant and reliable method for determining the MCE is by {\it directly} measuring $\Delta T_{ad}(T,\Delta B)$ under quasi-adiabatic conditions. The procedure comprises a full magnetization-demagnetization cycle, during which the experimental $T$ and $B$ are continuously recorded. In a half cycle, starting with the sample at an initial $T_i$, we magnetize (demagnetize) by gradually increasing (decreasing) the applied field from $B_i$ to $B_f$ and let the sample relax to the final $T_f$. In order to compute the temperature evolution for an {\it ideal} adiabatic process, one requires a precise knowledge of the heat that unavoidably is absorbed from (released to) the thermal bath during the direct measurement. For this purpose, the thermal conductance $k$ of the wires holding our sensor was previously determined as a function of $T$, using a standard copper piece as the sample. The non-adiabaticity induces a variation of the entropy $\Delta S=S(t)-S(t_0)$ in a time interval $t-t_0$, which can be expressed as $\Delta S=\int_{t_0}^{t}k(T_i-T)/T{\rm d}t$ at every time instant. Besides, from Eq.~\ref{eqS} we also have $\Delta S=\int_{T_{ad}}^{T}C(T,B)/T{\rm d}T$, where the adiabatic temperature $T_{ad}$, viz., the temperature if the sample would have been kept thermally isolated, is the only unknown and can therefore be deduced numerically. In our treatment, we safely disregard the entropy contribution due to the heat transferred from the sample holder to the refrigerant material, i.e., $\Delta S_{sh}=\int_{T_i}^{T_f} C_{sh}/T{\rm d}T$, since the heat capacity of the sample holder $C_{sh}$ is negligible with respect to that of Gd(OOCH)$_3$ below liquid-helium temperature.

Figure~4 shows the time evolution of $B$, $T$ and $T_{ad}$ for a full magnetization-demagnetization cycle, starting at $T_i=0.95$~K and for a field change~$\Delta B=B_f-B_i=(1-0)$~T or $(0-1)$~T, depending on whether we deal with the magnetization or demagnetization process, respectively. We note that the exact same conditions are highlighted in Figure~2: process ${\rm A}\rightarrow{\rm C}$ for the magnetization and process ${\rm D}\rightarrow{\rm B}$ for the demagnetization. In the top panel of Fig.~4, we observe $T$ to increase while we magnetize to 1~T. Here $T_{ad}$ increases more than $T$ because the thermal losses to the bath are compensated to obtain $T_{ad}$. Upon reaching $B_f$, $T$ decays back towards $T_i=0.95$~K but $T_{ad}=3.5$~K is constant, since it corresponds to an adiabatic process at constant $B$. In the bottom panel, $T$ decreases below $T_i$, while we demagnetize to zero field, whereupon $T$ gradually relaxes back to equilibrium, while constant $T_{ad}=0.45$~K. Remarkably, the final adiabatic temperatures of 3.5~K and 0.45~K obtained after sweeping the field up and down, respectively, corroborate the results independently inferred by an indirect method -- see states C and B, respectively, in Figure~2.

The MCE of Gd(OOCH)$_3$ is exceptionally large, especially in comparison with other molecule-based magnetic refrigerants. A fine example is the recently studied dimeric light molecule $[\{{\rm Gd}({\rm OAc})_3({\rm H}_2{\rm O})_2\}_2]\cdot 4{\rm H}_2{\rm O}$, hereafter shortened as $\{{\rm Gd}_2\}$, whose $-\Delta S_{m}$ reaches a value as large as 40.6~J~kg$^{-1}$~K$^{-1}$ for $\Delta B=(7-0)$~T and $T\simeq 1.8$~K.~\cite{Evangelisti11} For widespread applications, the interest is chiefly restricted to applied fields which can be produced with permanent magnets, viz., in the range $1-2$~T. In this regard $\{{\rm Gd}_2\}$ could be appealing because a weak, ferromagnetic intracluster exchange interaction enhances the MCE for low applied fields, yielding an outstanding $-\Delta S_{m}=27.0$~J~kg$^{-1}$~K$^{-1}$ for $\Delta B=(1-0)$~T and $T\simeq 0.8$~K. However, these values are not as large as the ones we here report for Gd(OOCH)$_3$, e.g., $-\Delta S_{m}=56.0$~J~kg$^{-1}$~K$^{-1}$ and 31.2~J~kg$^{-1}$~K$^{-1}$ for $\Delta B=(7-0)$~T and $(1-0)$~T, respectively, at similar corresponding temperatures such as seen in Figure~3.

Gadolinium gallium garnet (GGG) is {\it the} reference magnetic refrigerant material for the liquid-helium temperature region.~\cite{Daudin82,Numazawa03} Indeed, its functionality is commercially exploited in spite of a relatively modest maximum $-\Delta S_{m}=20.5$~J~kg$^{-1}$~K$^{-1}$ for $\Delta B=(2-0)$~T. This apparent contradiction is resolved by measuring the entropy change in terms of equivalent volumetric units, which take into consideration the GGG mass density $\rho=7.08$~g~cm$^{-3}$. By so-doing, GGG achieves a record value $-\rho\Delta S_{m}\simeq 145$~mJ~cm$^{-3}$~K$^{-1}$ for the same applied field change of 2~T. Although these units are not often used to characterize the MCE, they are better suited for assessing the implementation of the refrigerant material in a designed apparatus.~\cite{Gschneidner05} On this point, one could correctly argue that the MCE of molecule-based refrigerant materials is disfavored by their typically low $\rho$. For instance, $\rho=2.04$~g~cm$^{-3}$ for the aforementioned $\{{\rm Gd}_2\}$, which results in $-\rho\Delta S_{m}\simeq 56$~mJ~cm$^{-3}$~K$^{-1}$ and 73~mJ~cm$^{-3}$~K$^{-1}$ for $\Delta B=(1-0)$~T and $(3-0)$~T, respectively,~\cite{Evangelisti11} i.e., definitely much lower than GGG.

The mass density $\rho=3.86$~g~cm$^{-3}$ of Gd(OOCH)$_3$ is very large among molecule-based materials, though yet smaller than that of GGG. In Gd(OOCH)$_3$, the Gd$^{3+}$ centers are interconnected only by short and extremely light CHOO$^{-}$ ligands. Ultimately, this enhances the MCE favored by a larger weight of magnetic elements with respect to non-magnetic ones, which act passively. As a matter of fact, the mass density of these two materials is effectively counterbalanced by the magnetic:non-magnetic weight ratio, which amounts to 0.54 in Gd(OOCH)$_3$ and to a lower 0.47 in GGG.~\cite{note2} For comparison, this number further reduces to 0.39 in the case of $\{{\rm Gd}_2\}$. Overall, adopting the proper units, the MCE of Gd(OOCH)$_3$ is characterized by maxima $-\rho\Delta S_{m}\simeq 120$~mJ~cm$^{-3}$~K$^{-1}$ and 189~mJ~cm$^{-3}$~K$^{-1}$ for $\Delta B=(1-0)$~T and $(3-0)$~T, respectively, as can be seen in Figure~3. These values compares favorably with the ones obtained from GGG.

Concluding, we experimentally determine the magnetocaloric effect of the Gd(OOCH)$_3$ metal-organic framework material. Under quasi-adiabatic conditions, sub-Kelvin direct measurements of the temperature change corroborate the results inferred from indirect methods. By comparing the MCE per volume of other known materials, such as GGG, we demonstrate that gadolinium formate could serve as an excellent magnetic refrigerant for liquid helium temperatures. Our observations are interpreted as the result of a light and compact structural framework promoting very weak magnetic correlations between the Gd$^{3+}$ spin centers. Finally, we foresee that synthetic and technological strategies, already developed for the surface deposition of MOF materials, could ultimately facilitate the integration and exploitation of Gd(OOCH)$_3$ within molecule-based microdevices for on-chip local refrigeration.~\cite{Lorusso13}

\begin{acknowledgments}
We thank E. Moreno Pineda. This work has been supported by the Spanish MINECO through grant MAT2012-38318, and by an EU Marie Curie IEF (to G. L.).
\end{acknowledgments}


\begin{thebibliography}{}
\bibitem{note} We refrain from discussing the magnetic ordering mechanism any further since this topic exceeds the scope of this article and its full comprehension requires additional measurements, whose results will be exhaustively presented elsewhere.
\bibitem{note2} For Gd(OOCH)$_3$, the relative molecular mass and number of Gd$^{3+}$ ions per formula unit are $m=292.30$~g~mol$^{-1}$ and $n=1$, respectively. For GGG, $m=1012.37$~g~mol$^{-1}$ and $n=3$. For $\{{\rm Gd}_2\}$, $m=812.89$~g~mol$^{-1}$ and $n=2$. Therefore, $nA_{r}/m=0.54$, 0.47 and 0.39 for Gd(OOCH)$_3$, GGG and $\{{\rm Gd}_2\}$, respectively, where $A_r=157.25$~g~mol$^{-1}$ is the gadolinium relative atomic mass.
\end{thebibliography}
\end{document}